\documentstyle [12pt] {article}

\title{
 Optimization of crystal extraction experiment
}

\author{Valery Biryukov\thanks{E-mail:  biryukov@mx.ihep.su  }
\\ IHEP Protvino, 142284 Moscow Region, Russia }

\date{}
\begin{document}
\normalsize

\maketitle

\thispagestyle{empty}

\begin{abstract}
\normalsize
Using a computer model for the crystal extraction,
we investigate the bent-crystal
parameters optimal for the extraction experiment.
The optimal crystal curvature is found to be near 1 GeV/cm (for $pv/R$),
i.e. a factor of 2--3 higher than for
the crystal application in beam lines.
An influence of the accelerator optics on extraction is discussed.
A possibility of using the high-Z crystals for extraction is considered.
The simulations for the ongoing experiments at the CERN-SPS
and the Fermilab Tevatron, and for the proposed extraction at LHC,
 are presented.
\end{abstract}

\setcounter{page}{0}
\thispagestyle{empty}
\newpage
\normalsize

\section{Optimization of single-pass channeling}

Optimization of channeling in a single pass is well understood\cite{ufn}.
Two effects cause the particle dechanneling in a bent crystal:
the bending dechanneling due to centrifugal force,
and the scattering.
For a given bending angle $\Theta_B$,
a larger crystal length $L$ means  smaller curvature 1/$R$=$\Theta_B/L$
and hence smaller bending dechanneling.
However, the $L$ increase means larger loss for scattering
(characterized by the 1/$e$-dechanneling length $L_D$).
There is an optimal $L$, when losses from bending and scattering
are roughly equal.
In a harmonic approximation for the interplanar potential,
the bent-crystal transmission
can be estimated as
\begin{equation}	\label{effh}
F (\Theta_B, \rho)
      =  A_{S} \left(1- \rho\right)^2
	 \exp \left(-\frac{\Theta_B}{\Theta_D\rho
	 (1-\rho)^2}\right)  \; ,
\end{equation}
We used the notation $\rho$=$R_c/R$ ($R_c$ is the critical radius).
$A_S$ is the number of particles trapped in a straight crystal.
The factor $(1-R_c/R)^2$ describes the $A_S$ reduction due to
bending dechanneling.
The exponent describes a reduction due to scattering;
$\Theta_D$=$L_D/R_c$ is a constant defined by the crystal properties.
In a bent crystal, $L_D$ is reduced by $(1-R_c/R)^2$,
just the same factor as for bending dechanneling.
See review\cite{ufn} for further details and references.
The dislocation dechanneling
in crystals is reviewed in Ref.~\cite{pre2}.

\section{Optimization of multi-pass channeling}

\subsection{Qualitative discussion}

If a crystal works in the extraction mode
at a circular accelerator,
the circulating particles may pass through it many times.
The simplest estimate of the total (with multipasses) efficiency
of the particle extraction $F$ may be derived 
using the following two simplifications:

 (a) the probability $A_k$ of the particle capture into the channeling mode
 in the $k$-th passage through the crystal is the same as
 that probability in the first passage, $A_k=A_1=A$
 (actually $A_k$ decreases with $k$ as the
 beam divergence grows with the scattering in crystal);

 (b) the probability $q_k$ of the particle loss, either in the crystal
 or at the vacuum chamber wall,
 with the $k$-th passage through the crystal is the same as
 that probability in the first passage, $q_k=q_1=q$
 (actually $q_k$ may increase with $k$
 if the particles reach the vacuum chamber wall with scattering).

 The circulating particles are
 gradually removed from the ring with these two processes --
 crystal channeling and nuclear interaction,
 with respective probabilities $A$ and $q$ per a single passage of crystal.
The number of the particles remaining in the ring after $k$ passages
through the crystal is
$N_k = N_0 (1-A)^k (1-q)^k$
if the initial number was $N_0$.
The particle number extracted by the $(k+1)$-th passage equals
$F_kN_0 = A N_0 (1-A)^k (1-q)^k$,
and total efficiency amounts to
\begin{equation}	\label{e5}
F = \sum F_k
  = \frac{A}{A \cdot (1- q)+ q}
        \cdot \exp (-L/L_D)
\end{equation}
 The exponent takes into account dechanneling
 of the trap\-ped particles.
 Since both assumptions (a,b) tend to overestimate $F$,
 Eq.~(\ref{e5}) is the upper limit for
 the real efficiency.

 Eq.~(\ref{e5}) well agrees, e.g., with the
 computer simulation\cite{tar91} of proton extraction from
 a 20-TeV SuperCollider, Fig.~1.
 The analytical curve (\ref{e5}) was
 predicted in Ref.~\cite{bi91} for 90-$\mu$rad bending;
 the points are from the more-recent simulation
 for 100-$\mu$rad bending\cite{tar91}.

\begin{figure}[bth]
\begin{center}
\setlength{\unitlength}{1mm}\thicklines
\begin{picture}(85,75)(0,35)

 \put(3,85){\circle{2}}
 \put(6,90){ \circle{2}}
 \put(12,93){\circle{2}}
 \put(24,92){\circle{2}}
 \put(40,88){\circle{2}}
 \put(60,86){\circle{2}}
 \put(5,60){\small $\otimes$}
 \put(11,75){\small $\otimes$}
 \put(23,79){\small $\otimes$}
 \put(39,79){\small $\otimes$}
 \put(59,79){\small $\otimes$}

\put(2.2,76.7){\circle*{1}}
\put(2.4,82.2){\circle*{1}}
\put(2.6,85.2){\circle*{1}}
\put(3.0,88.2){\circle*{1}}
\put(4.0,91.) {\circle*{1}}
\put(5.0,92.) {\circle*{1}}
\put(6.0,92.4){\circle*{1}}
\put(7.0,92.5){\circle*{1}}
\put(8.0,92.6){\circle*{1}}
\put(9,92.5){\circle*{1}}
\put(10.,92.4){\circle*{1}}
\put(11.,92.3){\circle*{1}}
\put(12.,92.1){\circle*{1}}
\put(13.,91.9){\circle*{1}}
\put(14.,91.8){\circle*{1}}
\multiput(16,91.4)(2,-.4){2}{\circle*{1}}
\multiput(20,90.6)(2,-.45){2}{\circle*{1}}
\multiput(24,89.7)(2,-.4){2}{\circle*{1}}
\multiput(28,88.9)(2,-.35){2}{\circle*{1}}
\multiput(32,88.2)(2,-.4){2}{\circle*{1}}
\multiput(36,87.4)(2,-.35){6}{\circle*{1}}
\multiput(48,85.3)(2,-.325){4}{\circle*{1}}
\multiput(56,84.0)(2,-.275){8}{\circle*{1}}
\multiput(72,81.8)(2,-.25){4}{\circle*{1}}

\put(0,105) {\line(1,0){90}}
\put(0,40)  {\line(0,1){65}}
\put(90,40){\line(0,1){65}}
\put(0,40)  {\line(1,0){90}}
\multiput(4,40)(4,0){22}{\line(0,1){1.}}
\multiput(20,40)(20,0){4}{\line(0,1){2.}}
\multiput(4,105)(4,0){22}{\line(0,-1){1.}}
\multiput(20,105)(20,0){4}{\line(0,-1){2.}}
\put(19,42.5){\makebox(2,1)[b]{5}}
\put(39,42.5){\makebox(2,1)[b]{10}}
\put(59,42.5){\makebox(2,1)[b]{15}}
\put(79,42.5){\makebox(2,1)[b]{20}}
\multiput(0,45)(0,5){12}{\line(-1,0){1.}}
\multiput(0,50)(0,10){6}{\line(-1,0){2.}}
\multiput(90,45)(0,5){12}{\line(1,0){1.}}
\multiput(90,50)(0,10){6}{\line(1,0){2.}}
\put(-7,50){\makebox(2,1)[l]{50}}
\put(-7,60){\makebox(2,1)[l]{60}}
\put(-7,70){\makebox(2,1)[l]{70}}
\put(-7,80){\makebox(2,1)[l]{80}}
\put(-7,90){\makebox(2,1)[l]{90}}
\put(-9,100){\makebox(2,1)[l]{100}}

\put(-6,107){F (\%)}
\put(45,35){Crystal length (cm)}

\end{picture}
\end{center}
\caption {
 Extraction efficiency for a 20-TeV collider
as a function of the Si(110) crystal length.
The curve ($\bullet$) is analytical prediction Eq.~(2).
The other data are from computer simulation[4]:
the first-pass ($\otimes$) and the overall (o) efficiency.
 }		\label{ssc}
\end{figure}
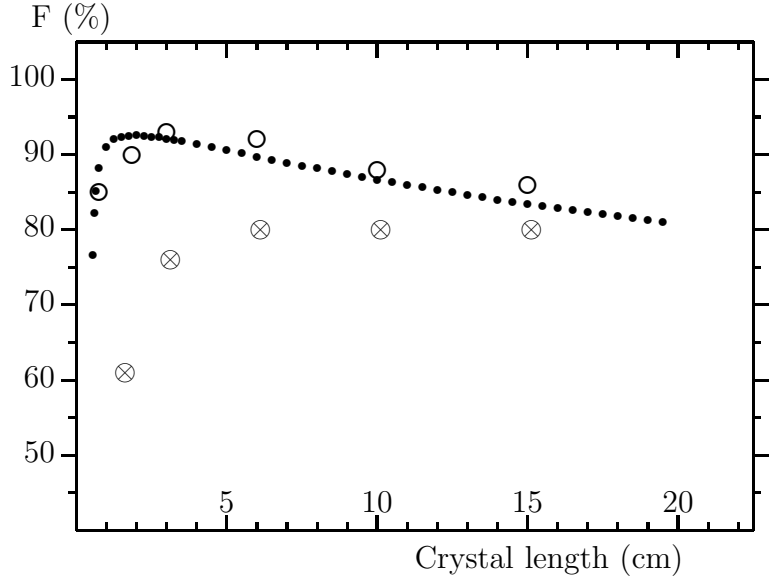

To understand Eq.~(\ref{e5}) better,
it is useful to consider several limiting cases.
\begin{enumerate}
\item
 If the probability $q$ of the unchanneled particle interaction
 with nuclei over the crystal length is very high, $q\simeq$1,
 the secondary passes are not possible.
 Eq.~(\ref{e5}) is reduced then to the efficiency of a single pass:
 $F\simeq A \exp (-L/L_D)$.
\item
Typically, $q\simeq L/L_N\simeq$0.1 over a single pass in crystal,
where $L_N$ is the nuclear interaction length.
If the probability of capture into channeling mode is very small,
$A\ll q$, Eq.~(\ref{e5}) is reduced to
 $F\simeq (1/q) A \exp (-L/L_D)$. This is simply
the single-pass efficiency times the mean number of passes $(1/q)$
before a nuclear interaction.
\item
The opposite case $q\ll A$ would mean that the circulating particles
are removed from the ring mostly with the crystal channeling
rather than with nuclear interactions.
The efficiency is then close to 1,
with Eq.(\ref{e5}) reduced to
 $F$$\simeq (1-q/A)\exp (-L/L_D)$.
\end{enumerate}

A shorter crystal disturbs the particles less.
The amplitude of betatron oscillation of a scattered proton
grows with the scattering angle squared $\theta_s^2$:
$x_{max}$=$(x_0^2+\beta^2\theta_s^2)^{1/2}$;
$\beta$ is the accelerator function, $x_0$ the crystal edge position.
When $x_{max}$ exceeds the aperture, the proton is lost.
Keeping $\theta_s^2$ below some critical value, one may
reduce $L$ to get a respective increase in the allowed
number of passes $\sim$1/$L$ before a loss.
Furthermore, the scattering over $L$ is reduced,
so the scattered protons have
smaller divergence in further passes.

On the other hand, a short crystal need to be more strongly bent to keep
$\Theta_B$. With $pv/R$ greater than
$\sim$1 GeV/cm in silicon, the crystal acceptance rapidly decreases.
This cuts the extraction efficiency at too-small $L$.

\subsection{Optimization of crystal parameters}

Detailed computer study of extraction
should take into account the accelerator optics (betatron motion,
aperture) and the processes
of channeling, scattering, and nuclear interaction in crystal.
These processes depend essentially on $L$.

\begin{figure}[htb]
\begin{center}
\setlength{\unitlength}{1.mm}
\begin{picture}(80,69)(-10,-3)
\thicklines

\put(3,20){\circle{2}}
\put(4,39){\circle{2}}
\put(5,54){\circle{2}}
\put(10,64){\circle{2}}
\put(20,62){\circle{2}}
\put(30,54){\circle{2}}
\put(40,52){\circle{2}}

\put(3,12){\circle*{2}}
\put(4,25){\circle*{2}}
\put(5,37){\circle*{2}}
\put(7,42){\circle*{2}}
\put(10,36){\circle*{2}}
\put(20,31){\circle*{2}}
\put(30,21){\circle*{2}}
\put(40,16){\circle*{2}}

\put(38.4,11.4){\small $\otimes$}
\put(8.4,28.0){\small $\otimes$}
\put(40,10){\line(0,1){5}}
\put(39,15){\line(1,0){2}}
\put(39,10){\line(1,0){2}}

\put(0,0) {\line(1,0){50}}
\put(0,0) {\line(0,1){70}}
\put(0,70) {\line(1,0){50}}
\put(50,0){\line(0,1){70}}
\multiput(10,0)(10,0){5}{\line(0,1){1}}
\put(9.5,2.){\makebox(1,.5)[b]{1}}
\put(19.5,2.){\makebox(1,.5)[b]{2}}
\put(29.5,2.){\makebox(1,.5)[b]{3}}
\put(39.5,2.){\makebox(1,.5)[b]{4}}
\multiput(0,10)(0,10){7}{\line(-1,0){1}}
\put(-8,20){\makebox(1,.5)[l]{0.2}}
\put(-8,40){\makebox(1,.5)[l]{0.4}}
\put(-8,60){\makebox(1,.5)[l]{0.6}}

\put(-7,68){ F}
\put(32,-5){ L (cm)}

\end{picture}
\caption{
The SPS extraction efficiency simulated
as a function of the Si(110) crystal length.
For perfect surface (o) and $t$=1 $\mu$m ($\bullet$).
The dots ($\otimes$) are for the U-shaped design and $t$=20 $\mu$m.
Also shown is the measured range of efficiencies, 10-15 \%
for the 4-cm U-shaped crystal.
}	\label{fl}
\end{center}
\end{figure}
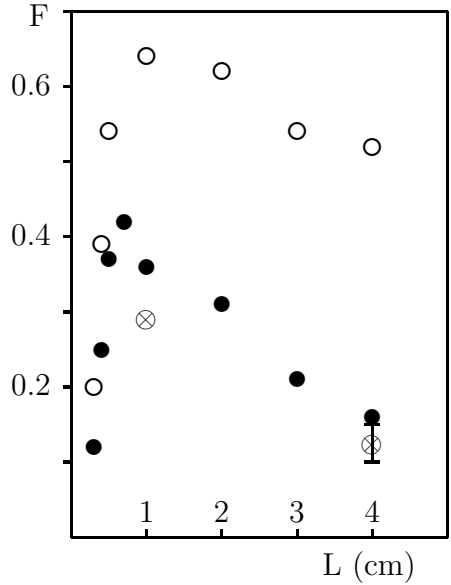

Figs.~2-4 show the extraction efficiency simulated as a function of $L$,
for the experiments at 120-GeV SPS \cite{sps,fer94,akb94,sps2},
900-GeV Tevatron \cite{car94,murphy}, and 7-TeV Large Hadron Collider
(see refs. in \cite{sps}).
The accelerator setting and the computer model have been earlier
described in Refs.\cite{bi78,tev,prl} respectively.
In Figs. 2-3 two cases were studied:
an ideal crystal, and the crystal with imperfect surface.
In the latter case an inefficient layer of some thickness $t$
appears near the crystal surface, defined by the surface irregularities.
In Fig.~4 we consider imperfect surface only.
Also shown in the SPS and LHC cases is the simulation for a realistic
model of the "U-shaped crystal" \cite{fer94,akb94,sps2} which included
the end straight parts (5 mm for $L$$\ge$5 cm and 1--4 mm for $L$$\le$4 cm),
and scattering in "the legs of U" and in the bending device.
The incident beam was nearly parallel at the SPS and LHC,
but divergent (11.5 $\mu$rad rms) at the Tevatron.
The efficiency measurements have been reported so far from the SPS
for the crystals of the 3--4 cm length only.
The simulation for the U-shaped crystal 4-cm long
in realistic model has given the efficiency in good agreement
with the experimental values of 10--15 \%, Fig.~2.
Notice that an agreement was found only in the assumption of
imperfect edge of crystal.
This edge imperfection (the first-pass suppression) predicted \cite{bi78}
for the SPS extraction was confirmed by the recent experiment \cite{sps2}.
Other comparisons of the expected and measured parameters
in crystal extraction see in Refs. \cite{sps,fer94,murphy}.

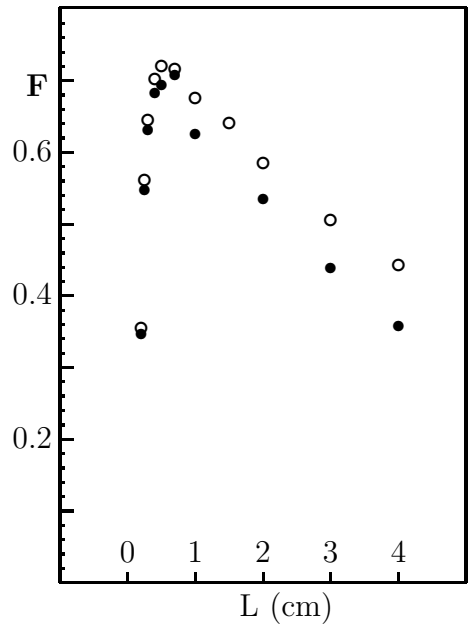
\begin{figure}[htb]
\begin{center}
\setlength{\unitlength}{.9mm}
\begin{picture}(80,83)(-20,-5)
\thicklines

\put(2,37.7){\circle{1.5}}
\put(2.5,59.6){\circle{1.5}}
\put(3,68.4){\circle{1.5}}
\put(4,74.5){\circle{1.5}}
\put(5,76.4){\circle{1.5}}
\put(7,76.0){\circle{1.5}}
\put(10,71.7){\circle{1.5}}
\put(15,67.9){\circle{1.5}}
\put(20,62.){\circle{1.5}}
\put(30,53.6){\circle{1.5}}
\put(40,47.0){\circle{1.5}}

\put(2,36.7){\circle*{1.5}}
\put(2.5,58.1){\circle*{1.5}}
\put(3,67.){\circle*{1.5}}
\put(4,72.4){\circle*{1.5}}
\put(5,73.6){\circle*{1.5}}
\put(7,75.0){\circle*{1.5}}
\put(10,66.3){\circle*{1.5}}
\put(20,56.7){\circle*{1.5}}
\put(30,46.5){\circle*{1.5}}
\put(40,38){\circle*{1.5}}

\put(-10,0) {\line(1,0){60}}
\put(-10,0) {\line(0,1){85}}
\put(-10,85) {\line(1,0){60}}
\put(50,0){\line(0,1){85}}
\multiput(0,0)(10,0){6}{\line(0,1){1.5}}
\put(-.5,3.){\makebox(1,.5)[b]{0}}
\put(9.5,3.){\makebox(1,.5)[b]{1}}
\put(19.5,3.){\makebox(1,.5)[b]{2}}
\put(29.5,3.){\makebox(1,.5)[b]{3}}
\put(39.5,3.){\makebox(1,.5)[b]{4}}
\multiput(-10,0)(0,10.6){8}{\line(1,0){2}}
\multiput(-10,0)(0,2.12){40}{\line(1,0){.75}}
\put(-17,21.2){\makebox(1,.5)[l]{0.2}}
\put(-17,42.4){\makebox(1,.5)[l]{0.4}}
\put(-17,63.6){\makebox(1,.5)[l]{0.6}}

\put(-15,72){\bf F}
\put(15,-5){ L (cm)}

\end{picture}
\caption {
The Tevatron extraction efficiency
for the ideal (o) and imperfect ($\bullet$), $t$=1 $\mu$m, crystals
as a function of the Si(110) crystal length.
      }	\label{fl2}
\end{center}
\end{figure}

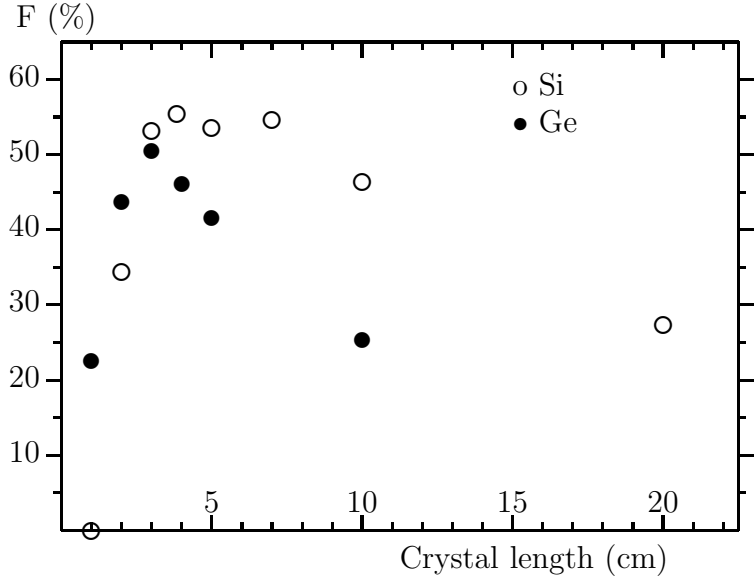
\begin{figure}[bth]
\begin{center}
\setlength{\unitlength}{1mm}\thicklines
\begin{picture}(85,75)(0,35)

 \put(4,40){\circle{2}}
 \put(8,74.4){\circle{2}}
 \put(12,93.1){\circle{2}}
 \put(14,95.4){ \circle{2}}
 \put(20,93.6){\circle{2}}
 \put(28,94.6){\circle{2}}
 \put(40,86.3){\circle{2}}
 \put(80,67.3){\circle{2}}

 \put(4,62.5){\circle*{2}}
 \put(8,83.7){\circle*{2}}
 \put(12,90.5){\circle*{2}}
 \put(16,86.1){\circle*{2}}
 \put(20,81.5){\circle*{2}}
 \put(40,65.3){\circle*{2}}

 \put(60,98){o Si}
 \put(60,93){$\bullet$ Ge}

\put(0,105) {\line(1,0){90}}
\put(0,40)  {\line(0,1){65}}
\put(90,40){\line(0,1){65}}
\put(0,40)  {\line(1,0){90}}
\multiput(4,40)(4,0){22}{\line(0,1){1.}}
\multiput(20,40)(20,0){4}{\line(0,1){2.}}
\multiput(4,105)(4,0){22}{\line(0,-1){1.}}
\multiput(20,105)(20,0){4}{\line(0,-1){2.}}
\put(19,42.5){\makebox(2,1)[b]{5}}
\put(39,42.5){\makebox(2,1)[b]{10}}
\put(59,42.5){\makebox(2,1)[b]{15}}
\put(79,42.5){\makebox(2,1)[b]{20}}
\multiput(0,45)(0,5){12}{\line(-1,0){1.}}
\multiput(0,50)(0,10){6}{\line(-1,0){2.}}
\multiput(90,45)(0,5){12}{\line(1,0){1.}}
\multiput(90,50)(0,10){6}{\line(1,0){2.}}
\put(-7,50){\makebox(2,1)[l]{10}}
\put(-7,60){\makebox(2,1)[l]{20}}
\put(-7,70){\makebox(2,1)[l]{30}}
\put(-7,80){\makebox(2,1)[l]{40}}
\put(-7,90){\makebox(2,1)[l]{50}}
\put(-7,100){\makebox(2,1)[l]{60}}

\put(-6,107){F (\%)}
\put(45,35){Crystal length (cm)}

\end{picture}
\end{center}
\caption {
Extraction efficiency for 7-TeV LHC
with Si(110) (o) and Ge(110) ($\bullet$)
"U-shaped" crystals (imperfect surface).
 }
\end{figure}

In all the Figs.~1--4, the optimum is found near $pv/R\simeq$1 GeV/cm.
The {\em short-crystal} optimum is most spectacular at lower energies,
$\sim$120 GeV, where scattering is appreciable.

All the circulating protons with sufficient amplitudes
are intercepted by the crystal sooner or later,
irrespective of its {\em transverse size}.
Therefore this parameter is not so important for extraction.

\subsection{Extraction with high--$Z$ crystals}

Crystals with higher $Z$ than in Si
(Ge and possibly W) have evident benefits,
such as higher critical angle $\theta_c$$\sim$$Z^{1/3}$,
and smaller $R_c$$\sim$1/$Z$; $\Theta_D$$\sim$$Z^{2/3}$.
Therefore the application of high-$Z$ perfect crystals
for a "single-pass" channeling is advantageous.
Interestingly, in crystal extraction
(which is multi-pass channeling essentially)
this may not be true.
The primary incident particles can be practically parallel
due to very small impact parameters $b$ (in the range of 1 $\mu$m)
at crystal; then $\theta_c$ doesn't matter.
In multipass mode, one has to trap the protons
{\em scattered} in the crystal.
Then the scattering angle and interaction length $L_N$
do matter as well as the channeling properties.

In general, one could consider a constant crystal lattice and vary $Z$ of
the atoms (we keep $Z/A$$\approx$const below).
Then $\theta_c$ is scaled as $Z^{1/3}$. The scattering
angle per unit length is scaled as $Z$. Increasing $Z$,
one scales the trapping efficiency as $Z^{-2/3}$
(i.e., the ratio of $\theta_c$ to particle divergence).
Since $R_c$$\sim$1/$Z$, one can shorten $L$$\sim$1/$Z$.
The scattering angle per a single passage of crystal is $\sim$$Z^{1/2}$,
and the efficiency is $Z^{-1/6}$ (decreasing with $Z$).

The final answer is more complicated, as one takes into account
the dechanneling loss (reduced in shorter crystal:
$L/L_D\sim Z^{-2/3}$), and absorption length $L_N$$\sim$$Z^{-2/3}$
(absorption loss is also reduced: $L/L_N\sim Z^{-1/3}$).
Fig.~4 shows the simulated efficiency of proton extraction from LHC
by Ge(110) crystal in the same conditions as for Si(110).
As expected, the efficiency with Ge crystal has slightly decreased
compared to Si.
Therefore, high-$Z$ crystals are hardly advantageous in extraction.

\subsection{Accelerator parameters}

The channeling efficiency is defined by the divergence of
incident particles w.r.t the crystal planes.
In the "diffusion mode" of extraction,
protons are intercepted by crystal with very small impact parameters $b$
in the first pass. Hand-in-hand, the primary divergence
in the "diffusion plane" is also small: $b'\ll\theta_c$.
The crystal may deflect protons either in this plane
(Protvino-CERN case\cite{bav89,sps}) or normally (Fermilab\cite{car94}).
In the latter case, primary divergence is scaled as $\beta^{-1/2}$.
Interestingly, in the Protvino-CERN case
the efficiency hardly depends on $\beta$.
The starting divergence of protons $b'\ll\theta_c$;
the secondary divergence is defined by the scattering
in crystal, which is independent of the machine optics.

Increasing $\beta$ one could increase primary $b$
to avoid a "septum width" $t$ due to crystal surface irregularities
(and get an efficient first-pass channeling).
However, recent measurements\cite{ches} indicate $t$$\simeq$50 $\mu$m,
while $b$ is well below 1 $\mu$m. In that case, the extraction becomes
insensitive to $\beta$.

The accelerator $\alpha$-function
sets a correlation of the mean incident angle $x'$ of
particles with the transverse coordinate:
$x'_{mean}$=$-\alpha x/\beta$.
The angle of incidence
varies by $\Delta x'$=$\alpha t_{cr}/\beta$
over the crystal thickness $t_{cr}$.
E.g., with $t_{cr}$=3.5 mm
in the SPS case, the $x'$ variation over the crystal face becomes
2.07$\times$3.5mm /90m$\approx$80 $\mu$rad $\gg\theta_c$=14 $\mu$rad!
A thick crystal ($t_{cr}\gg\beta\theta_c/\alpha$)
should loose in efficiency.

We leave aside the discussion of the betatron tune effects
in crystal extraction. Two points can however be mentioned.
At the rational values of tune the trajectory of a circulating
particle becomes periodical; if the first pass in crystal was
through an inefficient layer, some later passes suffer the same way
resulting in a dip of efficiency at the rational tunes \cite{bi78}.
Note however that in a real machine the other strong effects
are essential at the rational tunes.
Another relevant point is that crystal theorists predict \cite{vref,prl}
that in an aligned bent crystal the {\em unchanneled}
particles are deflected in average
at the angles of order $-\theta_c$,
opposite to the crystal bending.
In terms of the betatron motion,
a deflection at $\theta_c$ is a betatron phase advance
of $\Delta Q\approx\beta\theta_c/x_0$ per each traverse of crystal.
Typical $\Delta Q\simeq$0.1--0.2 is big and should be taken into account
in a study of the tune effects in crystal extraction.

\subsection{Pre-scattering}

To raise primary $b$,
some ideas were proposed\cite{tt1} and employed\cite{ass2}
to pre-scatter protons in a thin target before they hit a crystal.
In a target of length $s$,
the scattering angle is
$\theta_s$=$(E_s/E)(s/L_R)^{1/2}$; $E_s$=14 MeV, $L_R$ radiation length.
Then, the impact parameters in the first pass are
$b\simeq \beta^2\theta_s^2/2x_0$ \cite{tev},
which gives a relation for $s$:
\begin{equation}   \label{pre}
s \simeq
 \frac{2bx_0L_RE^2}{\beta^2E_s^2}
\end{equation}
Notice, $s\sim E^2$. E.g., requiring $b>$50 $\mu$m at crystal,
one needs a Si target as long as $s$$\ge$1 mm at 120-GeV SPS,
but $s$$\ge$40 mm at 900-GeV Tevatron.
For comparison, an effective length of
the first pass at SPS is  $\simeq$5 mm
(reduced near the {\em bent} face);
in the 900-GeV case, the 40 mm is as long as the crystal.
Therefore, an amorphous pre-scatter might work at $\sim$100 GeV and below.
At $\ge$1 TeV this idea can no longer be applied.
Notice that a pre-scatterer must be positioned
(closer to the beam than crystal)
with accuracy $\simeq$$b$ ($\simeq$50 $\mu$m)!
Otherwise, protons traverse it many times, and the benefits are lost.

The pre-scatter becomes more interesting with
a crystalline scat\-te\-rer\cite{tt1}, producing
angular kicks $\simeq$$\theta_c$.
The minimal length required is $\lambda$/4,
with $\lambda$ being the channeled-particle wavelength
($\lambda$$\simeq$0.1 mm at 1 TeV in Si).
The drawback is the need to tune two extra parameters:
position (with accuracy $\sim$10 $\mu$m) and angle
(with accuracy $<$$\theta_c$) of the pre-scatterer.

The alternative to the above ideas is the intrinsic scattering
in the edge of bent crystal.
It was shown with simulations for 120-GeV, 900-GeV
and 7-TeV extraction experiments that
a difference in efficiency between an ideal crystal and a crystal
with imperfect surface vanishes with increasing energy
$E$\cite{bi78,tev,prl}.
Therefore, the problem may disappear in TeV range.

\section{Conclusion}

The computer study of the crystal-extraction at accelerators indicates
 an interesting physics of "multipass channeling".
The optimal extraction is achieved with a {\em short} crystal,
and is due to the {\em multipass} mode
when the reduction in a single-pass transmission
is overcompensated with the increased mean number of passes.
The multipass mode may well be the only feasible one
at high energy accelerators, due to infinitesimal $b$.
This mode must be studied (at the optimum suggested) as it provides
a deeper insight into the crystal-extraction physics.
Different ideas for studying the multipass channeling
at circular accelerators have been reviewed in Refs.\cite{bi78,tev}.

\end{document}